\documentclass[11pt]{JHEP3} 

\usepackage{epsfig}
\usepackage{graphicx}
\usepackage{amsmath}
\usepackage{yfonts}
\usepackage{bm}
\usepackage{subfigure}

\def\be{\begin{equation}}\def\ba{\begin{eqnarray}}
\def\ee{\end{equation}}\def\ea{\end{eqnarray}}
\def\ben{\begin{enumerate}}\def\bitem{\begin{itemize}}
\def\een{\end{enumerate}}\def\eitem{\end{itemize}}
\def\no{\nonumber\\}

\def\calL{{\cal L}}

\newcommand{\e}{{\mbox{e}}}

\def\roughly#1{\mathrel{\raise.3ex\hbox{$#1$\kern-.75em%
\lower1ex\hbox{$\sim$}}}}

\def\deriv{\partial}

\def\<{\langle}
\def\>{\rangle}

\title{Holographic mesons in D4/D6 model revisited}

\author{Kwanghyun Jo
\\
Department of Physics Hanyang University, Seoul 133-791, Korea
\\ E-mail: \email{jokh38@gmail.com}}

\author{Youngman Kim
        \\
Asia Pacific Center
for Theoretical Physics and Department of Physics, Pohang University
of Science and Technology, Pohang, Gyeongbuk 790-784, Korea
\\  E-mail: \email{ykim@apctp.org}}
\author{Sang-Jin Sin
\\
Department of Physics Hanyang University, Seoul 133-791, Korea
\\ E-mail: \email{sjsin@hanyang.ac.kr}}
\received{\today}       
\preprint{ }
\abstract{
We revisit holographic mesons in  the D4/D6 model to study holographic light vector mesons and the properties of heavy quarkonium
in confining and deconfining phases. To treat the light mesons and heavy quarkonium on the same footing, we use the same compactification
scale $M_{KK}$ in both systems. We observe that like scalar and pseudo-scalar mesons the vector meson mass is
linearly proportional to the square root of the quark mass, when the quark mass is large.
With a $M_{KK}$ fixed by light meson masses, we calculate the mass of heavy quarkonium in confining and deconfining phases.
We also obtain the in-medium dispersion relation for heavy quarkonium, which is important to understand the dissociation and the screening mass
of heavy quarkonium in  quark-gluon plasma.
}

\begin{document}

\section{Introduction}

The properties of the heavy quarkonium both at zero and at finite
temperature have been intensively studied, see \cite{hQ_review} for a review.  At zero temperature, the
charmonium spectrum reflects detailed information about
confinement and quark-antiquark potentials in QCD~\cite{potential}.
At finite temperature, melting of heavy quarkonia could be a signal of the
formation of the quark-gluon plasma (QGP) in  a
relativistic heavy ion collision ~\cite{MS86}.
Moreover, lattice calculations suggest that the charmonium
states will survive at finite temperature up to about 1.6 to 2
times the critical temperature $T_c$\cite{AH04,Datta03}. This
suggests that analyzing the charmonium data from heavy ion
collision inevitably requires detailed information about the
properties of charmonium states in QGP. Therefore, it is very
important theoretical challenge to develop a consistent non-perturbative QCD picture for the heavy quark system both below and
above the QCD phase transition temperature.
In this respect, a promising attempt would be the holographic QCD (via AdS/CFT \cite{Maldacena}).

In a stringy D4/D6 model, scalar and pseudo-scalar bound states  have been studied
at zero and finite temperature in ~\cite{KruczD6, Mateos:2007vn}.
In a more phenomenological approach, bottom-up model, the mass spectrum of a charmonium
and its dissociation temperature have been investigated~\cite{KLL07, KKSS, FFMM}.
However, we note  that in the bottom-up model, different infrared scales are introduced to describe light mesons and heavy quarkonia.
For example, in the hard wall model, the location of the infrared cutoff $z_m$ varies from light mesons to heavy quarkonia:
$1/z_m\simeq 320~{\rm MeV}$ \cite{EKSS, PR} for the light meson and $1/z_m\simeq 1315~{\rm MeV}$ \cite{KLL07} for the charmonium.

In this paper we study the spectrum of the light vector meson and heavy quarkonium
 using the D4/D6 model~\cite{KruczD6} in high-temperature deconfining phase as well as in  confining phase.
  To treat the light mesons and heavy quarkonium on the same footing, we use the same compactification
scale $M_{KK}$ in both systems.
For the light vector meson, the spectrum was discussed in the D4/D8/$\bar D8$  model~\cite{SS} where
the chiral symmetry and its breaking are realized geometrically.
However, in the Sakai-Sugimoto model~\cite{SS} it is quite difficult to include the quark mass. Therefore it is still of worth to study the
spectrum of the vector mesons in the D4/D6 model  and study the
effect of the quark mass.
We observe that like scalar and pseudo-scalar mesons ~\cite{KruczD6} the vector meson mass is  proportional to a square root
of the quark mass $M_v^2 \sim m_q$ for large quark masses.
We also obtain the in-medium dispersion relation for heavy quarkonium, which is important to understand the dissociation and the screening mass
of heavy quarkonium~\cite{DKPW}.

\section{D4/D6 model}
We briefly summarize a pioneering holographic QCD model, the D4/D6 system~\cite{KruczD6}.
The  model contains N$_c$ number of D4 branes and N$_f$ flavor D6 branes whose configuration is given in Table 1.
\begin{table}[h]
\begin{center}
\begin{tabular}{|c|c|c|c|c|c|c|c|c|c|c|}
\hline
 &\multicolumn{4}{|c|}{\mbox{Boundary}} & $S^1$ & r( $S^2$)&\multicolumn{2}{|c|}{$S^2$}&\multicolumn{2}{|c|}{$D6_{\bot}$}\\
\hline
 &0&1&2&3& $\tau$ & $\lambda$ & $\psi_1$ &$\psi_2$&r&$\phi$\\
 \hline
D4&$\bullet$&$\bullet$&$\bullet$&$\bullet$&$\bullet$&&&&& \\
\hline
D6&$\bullet$&$\bullet$&$\bullet$&$\bullet$&&$\bullet$&$\bullet$&$\bullet$&& \\
\hline
\end{tabular}
\caption{The brane configuration : the background D4 and the probe D6 \label{braneprofile1}}
\end{center}
\end{table}
In the probe limit, the N$_c$ D4 branes are replaced by their supergravity background, and
 N$_f$ D6 branes are treated as probes. In this model mesons of a QCD-like gauge theory are described by
 the fluctuations of the D6 brane in the D4 background. The geometry of confining D4 brane reads
\ba
ds^2 &=& \left(\frac{U}{L}\right)^{3/2} (\eta_{\mu\nu}dx^\mu dx^\nu+f(U)d\tau^2)
+\left(\frac{L}{U}\right)^{3/2} \left(\frac{dU^2}{f(U)}+U^2 d\Omega_4^2\right) \no
\e^\phi &=& g_s \left(\frac{U}{L}\right)^{3/4}, \quad F_4 = dC_3 = \frac{2 \pi N_c}{V_4}\epsilon_4, \quad f(U) = 1 - \frac{U_K^3}{U^3}\, ,
\ea
where string coupling constant $g_s$ and the period of $\tau$  are given by
\be
g_s = \frac{g_{YM}^2}{2\pi M_{KK}l_s}, \quad \delta \tau \equiv \frac{4\pi}{3}\frac{L^{3/2}}{U_K^{1/2}}\, .
\ee
The parameter $L$ is given by the string coupling constant $g_s$ and the string length $l_s$, $L^3 = \pi g_s N_c l_s^3$,
and the compactification scale $M_{KK}$ reads
\be
 \quad M_{KK} = \frac{2\pi}{\delta \tau} = \frac{3}{2}\frac{U_K^{1/2}}{L^{3/2}}.
\ee
By introducing K($\rho$), the metric is simplified to be
\be
ds^2 = \left(\frac{U}{L}\right)^{3/2} \eta_{\mu\nu}dx^\mu dx^\nu + K(\rho)[d\lambda^2 + \lambda^2 d\Omega^2_2 + dr^2+r^2d\phi^2]\, ,
\ee
where
\be
K(\rho) \equiv L^{3/2}\frac{U^{1/2}}{\rho^2} , \quad  U(\rho) = \rho\left(1+\frac{U_K^3}{4 \rho^3}\right)^{2/3} \quad \mbox{where} ~ \rho^2 = \lambda^2 + r^2.
\ee
The position of the D6 brane is described by r($\lambda$) with $\phi$=0 and $\tau$ =constant. Then the induced metric on D6 is
\be
ds_{D6}^2=\left(\frac{U}{L}\right)^{3/2} \eta_{\mu\nu}dx^\mu dx^\nu + K(\rho)\Big[(1+\dot{r}^2)d \lambda^2 + \lambda^2 d\Omega_2^2\Big]\, .\label{inducedD6}
\ee
Now the action for D6 brane becomes
\be
S_{D6} = -T_6 \int d^7 \sigma \e^{-\phi} \sqrt{-\mbox{det} (g+2 \pi \alpha' F)}, \quad \mbox{where} \quad T_{D6} = \frac{1}{(2\pi)^6 l_s^7}\, .
\ee
 By using the well-known identity
\be
\det \left(
\begin{array}{cc}
A & B \\
C & D
\end{array}\right)
 = \det A \cdot \det (D-CA^{-1}B)
\ee
up to quadratic order in fields, we obtain
\be \label{DBIactionD6}
\calL_0 =- \frac{T_6}{g_s} \sqrt{h} \left(1+\frac{U_K^3}{4 \rho^3}\right)^2 \lambda^2 \sqrt{1+\dot{r}^2} \left[1+\frac{1}{4}(F_{\mu\nu}F^{\mu\nu}+2F_{\lambda\mu}F^{\lambda\mu})\right]
\ee
where $F_{\mu\nu} = \deriv_\mu A_\nu - \deriv_\nu A_\mu$.

\section{Meson spectroscopy}
In this section we are to compute the vector meson masses by considering gauge field fluctuations on the D6 brane.
Although we have two different scale of meson masses, light mesons and heavy quarkonia,
the origin of the difference is that of the quark mass rather than the interaction.
So we need to introduce only a single interaction scale $M_{KK}$.  This scale is to be  matched with the gauge theory scale
$\Lambda_{QCD}$, hence it is encoded in the background geometry while the
quark masses are  encoded in the geometry of the probe branes.

\subsection{Embedding}

We first find a D6 embedding geometry by solving the equation of motion for  r$(\lambda)$. From the DBI action (\ref{DBIactionD6}), we obtain the equation of motion for r($\lambda$) to be
\be
\deriv_\lambda\left[\left(1+\frac{U_K^3}{4\rho^3}\right)^2 \lambda^2 \frac{\dot{r}}{\sqrt{1+\dot{r}^2}}  \right] = -\frac{3}{2}\frac{U_K^3}{\rho^5} \left(1+\frac{U_K^3}{4\rho^3} \right) \lambda^2 r \sqrt{1+\dot{r}^2}.
\ee
With the following dimensionless variables
\be \label{dimensionlessrescale}
\lambda \rightarrow U_K \lambda, \quad r \rightarrow U_K r, \quad \rho \rightarrow U_K \rho\, ,
\ee
we rewrite it as
\be\label{D6embeddingeq1}
\deriv_\lambda\left[\left(1+\frac{1}{4\rho^3}\right)^2 \lambda^2 \frac{\dot{r}}{\sqrt{1+\dot{r}^2}}  \right] = -\frac{3}{2}\frac{1}{\rho^5} \left(1+\frac{1}{4\rho^3} \right) \lambda^2 r \sqrt{1+\dot{r}^2}.
\ee
For large $\lambda$, we can solve the equation of motion for $r(\lambda)$ to obtain the asymptotic solution as
\be
r(\lambda) \sim r_\infty + \frac{c}{\lambda}
\ee
where $r_\infty$, $c$ are related to the quark mass and the chiral condensate, see \cite{KruczD6} for details.

\subsection{Scalar and pseudo-scalar fluctuations}
We start with scalar and pseudo-scalar fluctuations.
Though these were extensively studied in \cite{KruczD6}, we include them for completeness, not to improve
 the results in  \cite{KruczD6}. The transverse fluctuation of the D6 brane is given by
\ba
r(x^\mu,\lambda)=r_v(\lambda)+\delta r(x^\mu),~ \phi(x^\mu,\lambda)=\delta \phi(x^\mu,\lambda)\, ,\label{scF}
\ea
where $r_v(\lambda)$ is the solution of the embedding equation.
Inserting (\ref{scF}) into the induced metric (\ref{inducedD6}) and the DBI action (\ref{DBIactionD6}), we obtain the induced metric
\ba
ds^2 &=& \left(\frac{U}{L}\right)^{3/2}\eta_{\mu\nu}dx^\mu dx^\nu + K [(1+\dot{r}_v^2)d\lambda^2+\lambda^2 d\Omega_2^2]+ 2K \dot{r}_v \deriv_a \delta r d\lambda dx^a \no
&&+ K[\deriv_a \delta r \deriv_b \delta r+ r_v^2 \deriv_a \delta \phi \deriv_b \delta \phi] dx^a dx^b\, ,
\ea
where $a$ and $b$ run over 0 to $\lambda$ and the DBI action, up to quadratic order,
\ba
\calL &=& \calL_0 - \frac{1}{2} T_{D6} U_K^3 \lambda^2 \sqrt{h}\sqrt{1+\dot{r}_v^2}\bigg[
\frac{U^3}{\rho_v^3(1+\dot{r}_v^2)}\left(\frac{(\deriv_\lambda \delta r)^2}{1+\dot{r}_v^2}+r_v^2 (\deriv_\lambda \delta \phi)^2 \right)+\frac{L^3 U^2}{U_K\rho_v^5} \left(\frac{\deriv_\mu \delta r\deriv^\mu \delta r}{1+\dot{r}_v^2} + r_v^2 \deriv_\mu \phi\deriv^\mu \phi\right) \no
&& -\frac{3}{2\rho_v^7}\left(\left(1+\frac{1}{4 \rho_v^3}\right)(\lambda^2 - 4 r_v^2) - \frac{3r_v^2}{4 \rho_v^3}\right) (\delta r)^2 - \frac{3r_v \dot{r}_v}{2\rho_v^5} \left(1+\frac{1}{4 \rho_v^3}\right) \frac{\deriv_\lambda (\delta r^2)}{1+\dot{r}_v^2}\bigg] .
\ea
Now we arrive at the linearized equation of motion for pseudo-scalar,
\be
0=\deriv_\lambda \left(\frac{\lambda^2 r_v^2}{\sqrt{1+\dot{r}_v^2}}\left(1+\frac{1}{4 \rho ^3}\right)^2 \deriv_\lambda \phi \right)+\frac{r_v^2 \lambda^2 \sqrt{1+\dot{r}_v^2}}{\rho_v^5} U^2 \frac{9 M_\phi^2}{4 M_K^2} \phi (\lambda)
\ee
and for scalar,
\ba
0 &=& \deriv_\lambda\left[\frac{\lambda^2}{(1+\dot{r}_v^2)^{3/2}} \left(1+\frac{1}{4 \rho ^3}\right)^2 \deriv_\lambda \delta r \right] + \frac{\lambda^2 U^2}{ \rho_v^5\sqrt{1+\dot{r}_v^2}} \frac{9 M_{\delta r}^2}{4 M_K^2} \delta r \\
&& +\frac{3 \lambda^2 \sqrt{1+\dot{r}_v^2}}{2 \rho_v^7} \left(\left(1+\frac{1}{4 \rho_v^3}\right)(\lambda^2 - 4 r_v^2) - \frac{3r_v^2}{4 \rho_v^3}\right)\delta r - \deriv_\lambda\left(\frac{3 \lambda^2}{2 \rho_v^5} \frac{r_v \dot{r}_v}{\sqrt{1+\dot{r}_v^2}}\left(1+\frac{1}{4 \rho_v^3}\right) \right) \delta r \nonumber.
\ea
These equations are numerically solved to get meson masses with proper boundary conditions~\cite{KruczD6}.

\subsection{Gauge field fluctuations}
Now we move on to the gauge field fluctuation.
The relevant part of the Lagrangian density for the gauge field is given by
\ba \label{gaugefieldaction}
\calL \sim -\frac{1}{4}\left(1+\frac{U_K^3}{4 \rho^3}\right)^2 \lambda^2 \sqrt{1+\dot{r}_v^2} \left(\frac{L}{U}\right)^{3/2}\left[\left(\frac{L}{U}\right)^{3/2} \eta^{\mu\nu} \eta^{\rho\sigma}F_{\mu\rho}F_{\nu\sigma}+2\frac{\eta^{\mu\nu}F_{\lambda\mu}F_{\nu \lambda}}{K(\rho)(1+\dot{r}_v^2)}\right]
\ea
where $r_v$ is the embedding solution.
We decompose the gauge fields in terms of the orthonormal basis $\psi_n, \phi_n$ as
\be
A_\mu(x^\mu,\lambda) = \sum_n B^{(n)}_\mu(x^\mu)\psi_n (\lambda), \quad
A_\lambda(x^\mu,\lambda) = \sum_n \varphi^{(n)}(x^\mu) \phi_n (\lambda)\, .
\ee
Then, the field strength takes the following form
\ba
F_{\mu\nu}(x^\mu,\lambda) &=& \sum_n F^{(n)}_{\mu\nu}(x^\mu) \psi_n(\lambda), \quad
F^{(n)}_{\mu\nu}(x^\mu) = \deriv_\mu B^{(n)}_\nu(x^\mu) - \deriv_\nu B^{(n)}_\mu(x^\mu) \, ,\no
F_{\mu\lambda}(x^\mu,\lambda)& =& \sum_n \bigg(\deriv_\mu \varphi^{(n)}(x) \phi_n(\lambda)-\deriv_\lambda\psi_n(\lambda) B^{(n)}_\mu\bigg)\, .
\ea
With the decomposition, the quadratic part of the Lagrangian for the $B_\mu$ field reads
\be \label{gaugeaction1}
\calL_{B} \sim -\frac{\sqrt{g_0}}{4} \sum_{m,n}\left[ \left(\frac{R}{U}\right)^{3/2}F^{(n)}_{\mu\nu} F^{(n)\mu\nu} \psi_m\psi_n +\frac{2}{K(\rho)(1+\dot{r}_v^2)}\dot{\psi}_m\dot{\psi}_n B_\mu B^\mu\right]\, ,
\ee
where $\sqrt{g_0}= \left(1+\frac{U_{KK}^3}{4 \rho^3}\right)^2 \lambda^2 \sqrt{1+\dot{r}_v^2} \left(\frac{L}{U}\right)^{3/2}$.
To recover the canonical kinetic term of the gauge field in 4D, we impose the normalization condition for
the wave function $\psi(\lambda)$ as
\be \label{psinorm}
(2\pi \alpha')^2 \tilde{T}_6 \int d \lambda \sqrt{g_0} \left(\frac{R}{U}\right)^{3/2} \psi_m\psi_n = \delta_{mn}
\ee
where $\tilde{T}_6 = -T_6 V_2/g_s$ and $V_2 = \int d \Omega_2 \sqrt{g_{66} g_{77}}$.
We will impose a similar condition for  $\phi (\lambda)$.
The wave function $\psi(\lambda)$ satisfies the following mode equation derived from the quadratic action
\be \label{eomforvectormode}
\deriv_\lambda \left(\sqrt{g_0}\frac{\deriv_\lambda \psi_n}{K(\rho)(1+\dot{r}_v^2)}\right) = -\sqrt{g_0}\left(\frac{L}{U}\right)^{3/2} m_n^2 \psi_n
\ee
where $m_n^2 = -q^2$.
Then, we obtain
\be \label{psieigenfunc}
(2\pi \alpha')^2 \tilde{T}_6 \int d \lambda \sqrt{g_0} \frac{1}{K(\rho)(1+\dot{r}_v^2)} \dot{\psi}_m \dot{\psi}_n = m_n^2 \delta_{mn}
\ee
where $m_n$ is the eigenvalue. From  (\ref{psinorm}) and (\ref{psieigenfunc}), we have now
\be
S_{D6} = -N \int d^4 x \sum_{n=1}^\infty\left(\frac{1}{4} F^{(m)}_{\mu\nu} F^{(n)\mu\nu} + \frac{1}{2} m_n^2 B^{(n)}_\mu B^{(n)\mu}\right)\, .
\ee
We rescale the coordinate by $U_K$  to obtain
\ba \label{eomforvectordimless}
\deriv_\lambda \left(\sqrt{g_0}\frac{\deriv_\lambda \psi_n}{K(\rho)(1+\dot{r}_v^2)}\right) = -\frac{\sqrt{g_0}}{U^{3/2}}\frac{9}{4}\frac{m_n^2}{M_{KK}^2} \psi_n\, .
\ea
To solve this equation, we impose two boundary conditions: at the IR, either $\psi_n(0)$ or $\dot{\psi}_n(0)=0$,
and at the UV, $\psi_n \sim \lambda^\alpha$ with $\alpha \leq$ 1/2 from the normalizability condition  in (\ref{psinorm}).

Now we consider $\phi_n$.
The normalization of $\phi_n$ is similar to (\ref{psieigenfunc})
\be
(2\pi \alpha')^2 \tilde{T}_6 \int d \lambda \sqrt{g_0} \frac{1}{K(\rho)(1+\dot{r}_v^2)} \phi_m \phi_n = \delta_{mn}\, .
\ee
As explained below and also in \cite{SS}, once
  we set the field as $\phi_n = \dot{\psi}_n/ m_n$ for n$\geq$1, it can be gauged away as a part of $B_\mu$ field.
However, zero mode is exceptional, which is orthogonal to the other modes
\be
(\phi_0,\phi_n) = \frac{(2\pi \alpha')^2}{m_n} \tilde{T}_6 \int d \lambda \sqrt{g_0} \frac{1}{K(\rho)(1+\dot{r}_v^2)} \phi_0 \dot{\psi}_n = 0 \quad (\mbox{for n} \geq 1)\, .
\ee
If we take $\phi_0 =C K(\rho)(1+\dot{r}_v^2)/\sqrt{g_0}$,
\be
(\phi_0,\phi_n) = \int^\infty_0 d \lambda ~\dot{\psi}_n = \psi_n(\infty)-\psi_n(0) = 0 \quad (\mbox{for n} \geq 1).
\ee
Then, the constant $C$ is given by
\be \label{cnorm}
1=(\phi_0,\phi_0) ~\rightarrow ~C= \left((2\pi \alpha')^2 \tilde{T}_6 \int d \lambda \frac{K(\rho)(1+\dot{r}_v^2)}{\sqrt{g_0}}\right)^{-1/2}.
\ee
The field strength is written as
\be
F_{\mu\lambda}(x^\mu,\lambda) = \deriv_\mu \varphi^0(x) \phi_0(\lambda)+
\sum_n \bigg(m_n^{-1}\deriv_\mu \varphi^{(n)}(x) - B^{(n)}_\mu \bigg) \dot{\psi}_n(\lambda)\, .
\ee
By gauge transformation, $B_\mu$ absorbs  $\deriv_\mu \varphi^{(n)}$,
\be
B^{(n)}_\mu \rightarrow B^{(n)}_\mu + m_n^{-1}\deriv_\mu \varphi^{(n)}(x)\, ,
\ee
and therefore the action  (\ref{gaugeaction1}) becomes
\be
S_{D6} = \int d^4 x \left[\frac{1}{2}\deriv_\mu \varphi^{0} \deriv^\mu \varphi^0+ \sum_{n=1}^\infty\left(\frac{1}{4} F^{(m)}_{\mu\nu} F^{(n)\mu\nu} + \frac{1}{2} m_n^2 B^{(n)}_\mu B^{(n)\mu}\right)\right].
\ee
Note that for the heavy quarkonium system,
\ba
C &=& \left((2\pi \alpha')^2 \tilde{T}_6 \int_0^\infty d \lambda \frac{\sqrt{1+\dot{r}_v^2}}{\lambda^2}\left(1+\frac{1}{4\rho^3}\right)^{-2/3} \right)^{-1/2} \no
&=& \left((2\pi \alpha')^2 \tilde{T}_6 \int_0^\infty d \lambda \frac{1}{\lambda^2}\right)^{-1/2} =\left((2\pi \alpha')^2 \tilde{T}_6 \frac{1}{\lambda}\bigg|^0_\infty \right)^{-1/2} =0\, ,
\ea
and so $\varphi_0$ =0 due to (\ref{cnorm}).
To impose the UV boundary condition for $\psi_n$ more precisely, we consider the mode equation (\ref{eomforvectordimless}) at large $\lambda$,
\ba
 \deriv_\lambda (\lambda^2 \deriv_\lambda \psi_n)=-\frac{m_n^2}{\lambda}\psi_n\, .
\ea
With $\psi_n \sim \lambda^\alpha$, we obtain $ \alpha(\alpha-3)=0.$
Since the normalizability condition dictates $\alpha\le1/2$, we should choose $\alpha$ = 0.

\subsection{Numerical results}
We solve the mode equations for scalar, pseudo-scalar, and gauge field fluctuations numerically. We first compute the mass of light mesons to fix the model parameter $r_\infty^l$ and $M_{KK}$. Since D4/D6 model has no non-Abelian chiral symmetry but for U(1)$_A$, the pseudo-scalar meson in this model corresponds to $\eta^\prime$ in QCD~\cite{KruczD6}. In QCD, however, U(1)$_A$ symmetry is explicitly broken by the axial anomaly, and the observed mass of $\eta^\prime$, $m_{\eta^\prime}=958~{\rm MeV}$, is much larger compared to the pion or kaon mass.
Note that some portion of the $\eta^\prime$ mass comes from the anomaly effect which scales as $N_f/N_c$.
Since we are working in the large $N_c$ limit, we may use the mass of $\eta^\prime$ with the anomaly contribution turned-off.
 So we use non-anomalous $\eta^\prime$ mass to obtain a rough number for the model input. To this end, we use the mass relation for Goldstone boson obtained in chiral perturbation theory at large $N_c$ \cite{EHS}:
\ba
m_\pi^2=\frac{2m_q\Sigma }{f_\pi^2}, \quad m_{\eta^\prime}^2=\frac{2\Sigma(2m_q+m_s)}{3f_\pi^2}+\frac{6\tau}{f_\pi^2}\, ,
\ea
where $m_u=m_d\equiv m_q$. The term with $\tau$ is from the axial anomaly.
Now we take $\tau=0$ to estimate the $\eta^\prime$ mass from non-anomalous contribution.
With $m_q=7~{\rm MeV}$, $m_s=150~{\rm MeV}$, $f_\pi=93~{\rm MeV}$, and $\Sigma=(230~{\rm MeV})^3$, we obtain
$m_\pi\sim 140~{\rm MeV}$ and $m_\eta^\prime \sim 390~{\rm MeV}$.
Note that the mass of $q\bar q$ bound state such as $\rho$ meson mass stays almost constant as we increase $N_c$: for instance the light meson mass at
large $N_c$  is extensively studied in a unitarized chiral perturbation theory~\cite{Pelaez} and also in lattice QCD~\cite{lattice12}.

\begin{figure}
\begin{center}
\includegraphics[width=0.5 \textwidth]{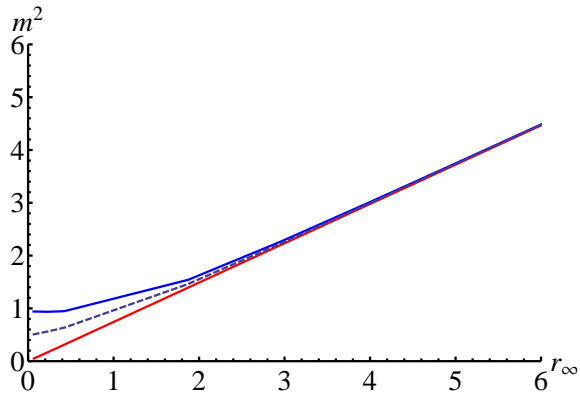}
\end{center}
  \caption{\small Meson masses for scalar (blue), pseudo-scalar (red), vector mode (dashed). The pseudo-scalar meson mass vanishes when current quark mass goes zero, see red line. For the high enough quark mass, all the meson masses are degenerated because of supersymmetry restoration \cite{KruczD6}. }\label{msvsmq}
\end{figure}

We use $\rho$-meson mass and $\eta^\prime$ mass in large $N_c$ limit as inputs to fix $r_\infty=0.191$ and $M_{KK}=1.039$.
Our fitting results are summarized in Table 2.
In Fig. \ref{msvsmq}, we plot the masses of scalar, pseudo-scalar and vector as a function of $r_\infty$.
As observed in  \cite{KruczD6}, for large  $r_\infty$
 the meson mass becomes degenerate and increases monotonically with $r_\infty$.
This is simply because the equation of motion for scalar, pseudo-scalar and vector for heavy quark system, $r_\infty \gg U_K$, are degenerated.
\be \label{degeneq}
\frac{L^3}{r_\infty} \frac{M^2}{(1+y^2)^{3/2}} \Psi + \frac{1}{y^2}\deriv_y(y^2 \deriv_y \Psi)=0\, ,
\ee
where $\Psi$ can be real scalar $\delta r$, pseudo-scalar $ \phi$, or vector $\psi$, and y is a
rescaled coordinate y=$\lambda/r_\infty$.
In the equation (\ref{degeneq}), the only scale is
\be
M^2 \sim \frac{L^3}{r_\infty} \sim \frac{m_q M_K}{\lambda_t}.
\ee
This means that for the heavy quark system, the fluctuating field has mass being proportional to $\sqrt{\frac{m_q M_K}{\lambda_t}}$.
As discussed in \cite{KruczD6}, the reason for the degeneracy is supersymmetry restoration. They are all in the same supermultiplet and for small $m_q$ limit supersymmetry is broken so their masses split.
 However, for the large quark mass or large separation between D4 and D6, the  embedding is nearly flat, and so D6 brane restores supersymmetry.

\begin{table}[h]
\begin{center}
\begin{tabular}{|c|c|cc|}
\hline
Mode& Input (MeV)& M/M$_{KK}$ & M (MeV) \\
\hline
Ps ~ ($\eta^\prime$)  & 390 &  0.375  &  390 \\
Rs ~ ($\sigma$)       &     &  0.918  &  954 \\
V ~ ($\rho$)          & 770 &  0.741  &  770 \\
\hline
\end{tabular}
\caption{Light meson masses to fix free parameters, $M_{KK}$ and $r_\infty$, in the model. }
\end{center}
\end{table}
Now we move on to the heavy quarkonium.
To describe the light meson and heavy quarkonium systems on the same footing,
we use the value of $M_{KK}$ determined in the light quark sector to calculate the mass of heavy quarkonium.
We choose $J/\psi$ mass as an input to fix $r_\infty^c$.
In Table 3, we list our results on the charmonium with $r_\infty^c = 11.92$ , and in Table 4 we show bottomonium masses
$r_\infty^b = 108.15$.
\begin{table}[h]
\begin{center}
\begin{tabular}{|c|c|ccc|}
\hline
Mode & Particle data & M/M$_{KK}$ & M(GeV) & Error\\
\hline
Ps ($\eta_c$) & 2.980 GeV & 2.978 & 3.095 & 3.72 \% \\
Re ($\chi$)   & 3.414 GeV & 2.979 & 3.096 & 9.31 \% \\
V  ($J/\psi$) & 3.096 GeV & 2.978 & 3.095 & ($\star$) \\
\hline
\end{tabular}
\caption{Charmonium mass. }
\end{center}
\end{table}
Note that $r_\infty$ is related to the quark mass parameter of D4/D6 model as\footnote{As well-known, the quark mass in D4/D6 model
could be different from that in QCD by a constant factor. To obtain the constant we need to compare the scalar two point function
obtained in D4/D6 model
with that in the operator product expansion of QCD, for example see~\cite{PR2}.}
\be \label{quarkmass}
m_q = \frac{U_K r_\infty}{2 \pi l_s^2} = \frac{r_\infty}{9\pi} g_{YM}^2 N_c M_K =\frac{r_\infty}{9\pi} M_K \lambda_t .
\ee
\begin{table}[h]
\begin{center}
\begin{tabular}{|c|c|ccc|}
\hline
Mode & Particle data  & M/M$_{KK}$ & M(GeV) & Error\\
\hline
Ps ($\eta_b$)   & 9.389 GeV  & 9.105 & 9.46 & 0.75 \% \\
Rs ($\chi$)     & 9.859 GeV  & 9.105 & 9.46 & 4.05 \% \\
V  ($\Upsilon$) & 9.460 GeV  & 9.105 & 9.46 & ($\star$) \\
\hline
\end{tabular}
\caption{Bottomonium mass. }
\end{center}
\end{table}
So far we have tried to study  light mesons and heavy quarkonia on the same footing, {\it i.e.} using the same value of $M_{KK}$ for
both cases. As long as the lowest lying KK mode masses are concerned, this unified approach seems working apart from the axial-vector meson.
Note that this defect of degenerate vector and axial-vector mesons is the feature of D4/D6 model itself not that of unified approach.
However, the way this attempted unified approach works
seems almost guaranteed in the following sense. The light meson masses are mostly inputs to fix
the $r_\infty^l$ and $M_{KK}$, while for heavy quarkonia $r_\infty^l$ will be fixed to reproduce
a heavy quarkonium mass.
As in Fig. (\ref{msvsmq}), the masses of heavy quarkonium ($r_\infty^h=10.72$) are almost degenerate regardless of their quantum numbers, while
experimentally those masses are not very different from each other.
Therefore, our results on the heavy quarkonium will be within experiments by roughly $10 \%$ deviations.

\section{Heavy quarkonium in deconfined phase}
We consider the meson spectroscopy in deconfining phase. By double Wick rotation, we get the black D4 background which is dual to the deconfining phase,
\ba
ds^2 &=& \left(\frac{U}{L}\right)^{3/2} (-f(U) dt^2 + d\vec{x}^2+d\tau^2)
+\left(\frac{R}{L}\right)^{3/2} \left(\frac{dU^2}{f(U)}+U^2 d\Omega_4^2\right) \no
\e^\phi &=& g_s \left(\frac{U}{L}\right)^{3/4}, \quad F_4 = dC_3 = \frac{2 \pi N_c}{V_4}\epsilon_4, \quad f(U) = 1 - \frac{U_T^3}{U^3}.
\ea
Here the temperature of the black D4 brane is identified with the temperature of a dual gauge theory
\be
\delta t_E = \frac{1}{T}, \quad \delta \tau = \frac{2 \pi}{M_{KK}}\, .
\ee
In terms of these periods, the metric parameters are given by
\be
U_K = \left(\frac{4 \pi}{3 \delta \tau}\right)^2 L^3, \quad U_T = \left(\frac{4 \pi}{3 \delta t_E}\right)^2 L^3\, .
\ee
As discussed in \cite{KruczD6}, there exists first order phase transition between confining and black D4 brane backgrounds; this phase transition is a deconfinement phase transition in field theory side. The critical temperature of the transition is fixed by the condition $\delta \tau = \delta t_E$
\be
T_{dec} = \frac{M_K}{2 \pi} = \frac{3}{4 \pi} \sqrt{\frac{U_K}{L^3}}.
\ee
Note that the temperature of the gauge theory is given as
\be\label{tdefineeq}
T = \frac{3}{4 \pi} \sqrt{\frac{U_T}{L^3}} = \frac{\bar{M}}{\sqrt{r_\infty}} ,\quad \mbox{where}~ \bar{M} = \sqrt{\frac{9}{4\pi}\frac{m_q M_{KK}}{\lambda_t}}.
\ee

\subsection{Embedding}
The D6 brane embedding in the black D4 background is described by r($\lambda$)
\be
ds^2 = \left(\frac{U}{L}\right)^{3/2} \bigg[-f(U)dt^2+d\vec{x}^2 \bigg] + K(\rho)\bigg[(1+\dot{r}^2)d\lambda^2 + \lambda^2 d\Omega_2^2 \bigg]
\ee
The DBI action for r($\lambda$) is then
\ba
S_{D6} &=& -\frac{1}{(2 \pi)^6 l_s^7} \int d^7\sigma \e^{-\phi} \sqrt{\mbox{det}g} \no
&=&-T_{D6} \int d^7\sigma \sqrt{h} \lambda^2 \sqrt{f(U) (1+\dot{r}^2)}\left(1+\frac{U_T^3}{4\rho^3}\right)^2\, .
\ea
{}From this, we obtain the equation of motion for the embedding function r($\lambda$)
\be
\deriv_\lambda \left[\left(1-\frac{U_T^6}{16 \rho^6}\right)\frac{\lambda^2 \dot{r}}{\sqrt{1+\dot{r}^2}}\right] - \frac{3 U_T^6 }{8 \rho^8}\sqrt{1+\dot{r}^2} \lambda^2 r=0.
\ee
The solution of the embedding equation was extensively studied in \cite{KruczD6}.
By comparing the energy density of various embedding solutions, the authors of \cite{KruczD6} found that there is  a first order phase transition at $r_\infty \sim U_K$. This transition, from Minkowski to black hole embedding, occurs when the D6 brane touches the black hole horizon at sufficiently large temperature.
 The transition temperature for charm quark is given by $T_{fund} \sim 1.0202 \bar{M}_c$,
 where $\bar{M}_c^2=\frac{r_\infty^c}{4\pi^2} M_{KK}^2 =(0.571 GeV)^2$.
 With the parameters fixed in confining phase, we obtain  $T_{dec} \sim$ 165 MeV and $T_{fund} \sim$ 582 MeV.
We summarize the different embedding solutions below.
\begin{table}[h]
\begin{center}
\begin{tabular}{|c|c|c|}
\hline
T & Embedding & Energy spectrum \\
\hline
$T < T_{dec}$                     & Confining emb. & discrete     \\
$T_{\mbox{dec}} < T < T_{fund}$   & Minkowski emb. & discrete     \\
$T_{fund} < T $                   & Blackhole emb. & continuous   \\
\hline
\end{tabular}
\caption{Classification of embeddings }
\end{center}
\end{table}

We note, however, that there is no priori reason for $T_{dec} < T_{fund}$, which is satisfied only when
\be \label{condforMik}
m_q^{f.s.} > \frac{0.98}{9 \pi} \lambda_t M_{KK} \simeq 36\lambda_t ~ \mbox{MeV}\, .
\ee
As a consequence, there is no Minkowski embedding for the light quark system, meaning that light mesons will be melted away immediately
when the temperature exceeds the deconfinement temperature $T>T_{dec}$.
For the heavy quark system we expect that there are discrete meson spectra even in the deconfining phase since
$m_q^h$ is big enough to satisfy the condition.

\subsection{Fluctuating fields as mesons}
In this section we study temperature dependent meson masses.
The temperature dependence of scalar and pseudo-scalar heavy quarkonia
are already considered in \cite{Mateos:2007vn}, and now we extend the work by including vector mesons.
In the high-temperature deconfining phase, there is only one relevant scale, $T/M$, which is nothing but the separation between D4 and D6 branes. Therefore varying temperature is equivalent to varying the D4-D6 separation with a fixed quark mass.

\subsubsection{Scalar and pseudo-scalar fluctuations}
Scalar and pseudo-scalar mesons are described by the D6 brane fluctuating on the transverse direction, $r(\lambda)=r_v(\lambda) + \delta r(t,\lambda)$, $\phi(t,\lambda)$. The quadratic part of Lagrangian after rescaling by $U_T$ is
\ba
\calL_{sc} & \sim & \frac{\sqrt{h}}{2 \sqrt{1+\dot{r}_v^2}} \lambda^2 \bigg[\left(1-\frac{1}{16 \rho_v^6}\right)\frac{(\deriv_\lambda \delta r)^2}{1+\dot{r}_v^2}  + \frac{3 r_v \dot{r}_v}{8 \rho_v^8} \deriv_\lambda(\delta r^2) +\bigg\{\frac{3(1+\dot{r}_v^2) \left(\lambda ^2-7r_v^2 \right)}{8 \rho_v^{10}} \\
&& - \left(1+\frac{1}{4\rho_v^3}\right)^{1/3} \frac{4}{1-4\rho_v^3} \left(\left(1+\frac{1}{4\rho_v^3}\right)^2\tilde{w}_r^2 - \left(1-\frac{1}{4\rho_v^3}\right)^2 \tilde{k}_r^2 \right) \bigg\} \delta r^2 + r_v^2\left(1-\frac{1}{16 \rho_v^6}\right) (\deriv_\lambda \phi)^2 \no
&& - \left(1+\frac{1}{4\rho_v^3}\right)^{1/3} \frac{4(1+\dot{r}_v^2) r_v^2}{1-4\rho_v^3} \left(\left(1+\frac{1}{4\rho_v^3}\right)^2\tilde{w}_\phi^2 - \left(1-\frac{1}{4\rho_v^3}\right)^2 \tilde{k}_\phi^2 \right) \phi^2 \bigg] \nonumber
\ea
The equation of motion for the bulk scalar field ($\delta r$) and for pseudo-scalar ($\phi$),
\ba
0 & = & \deriv_\lambda \left[\left(1-\frac{1}{16\rho_v^6}\right)\frac{\lambda^2 \deriv_\lambda \delta r}{(1+\dot{r}_v^2)^{3/2}}\right]-\left[\frac{3\lambda ^2(\lambda^2-7r_v^2)\sqrt{1+\dot{r}_v^2}}{8\rho_v^{10}} - \frac{3}{8}\deriv_\lambda\left(\frac{\lambda ^2}{\rho_v^8}\frac{r_v \dot{r}_v}{\sqrt{1+\dot{r}_v^2}}\right) \right]\delta r \no
&& + \left(1+\frac{1}{4\rho_v^3}\right)^{1/3} \frac{4 \lambda^2}{\sqrt{1+\dot{r}_v^2}(1-4\rho_v^3)} \left(\left(1+\frac{1}{4\rho_v^3}\right)^2\tilde{w}_r^2 - \left(1-\frac{1}{4\rho_v^3}\right)^2 \tilde{k}_r^2 \right) \bigg\}\delta r
\ea
where $ \tilde{w}^2 = \frac{L^3}{U_T} w , \tilde{k}^2 = \frac{L^3}{U_T} k^2 $,
 and the equation of motion for the pseudo-scalar is
\be
0 = \deriv_\lambda \left[\left(1-\frac{1}{16\rho_v^6}\right) \frac{\lambda^2 r_v^2}{\sqrt{1+\dot{r}_v^2}} \deriv_\lambda \phi\right] +\left(1+\frac{1}{4\rho_v^3}\right)^{1/3} \frac{4 \lambda^2 \sqrt{1+\dot{r}_v^2} r_v^2}{1-4\rho_v^3} \left(\left(1+\frac{1}{4\rho_v^3}\right)^2\tilde{w}_\phi^2 - \left(1-\frac{1}{4\rho_v^3}\right)^2 \tilde{k}_\phi^2 \right)\phi.
\ee
Since there is no Lorentz symmetry due to the temperature, we define the thermal meson mass as
 $\deriv_t^2 \phi =  m^2_\phi \phi$, {\it i.e.}, at vanishing spatial momentum.
\begin{figure}
\begin{center}
\includegraphics[width=0.42 \textwidth]{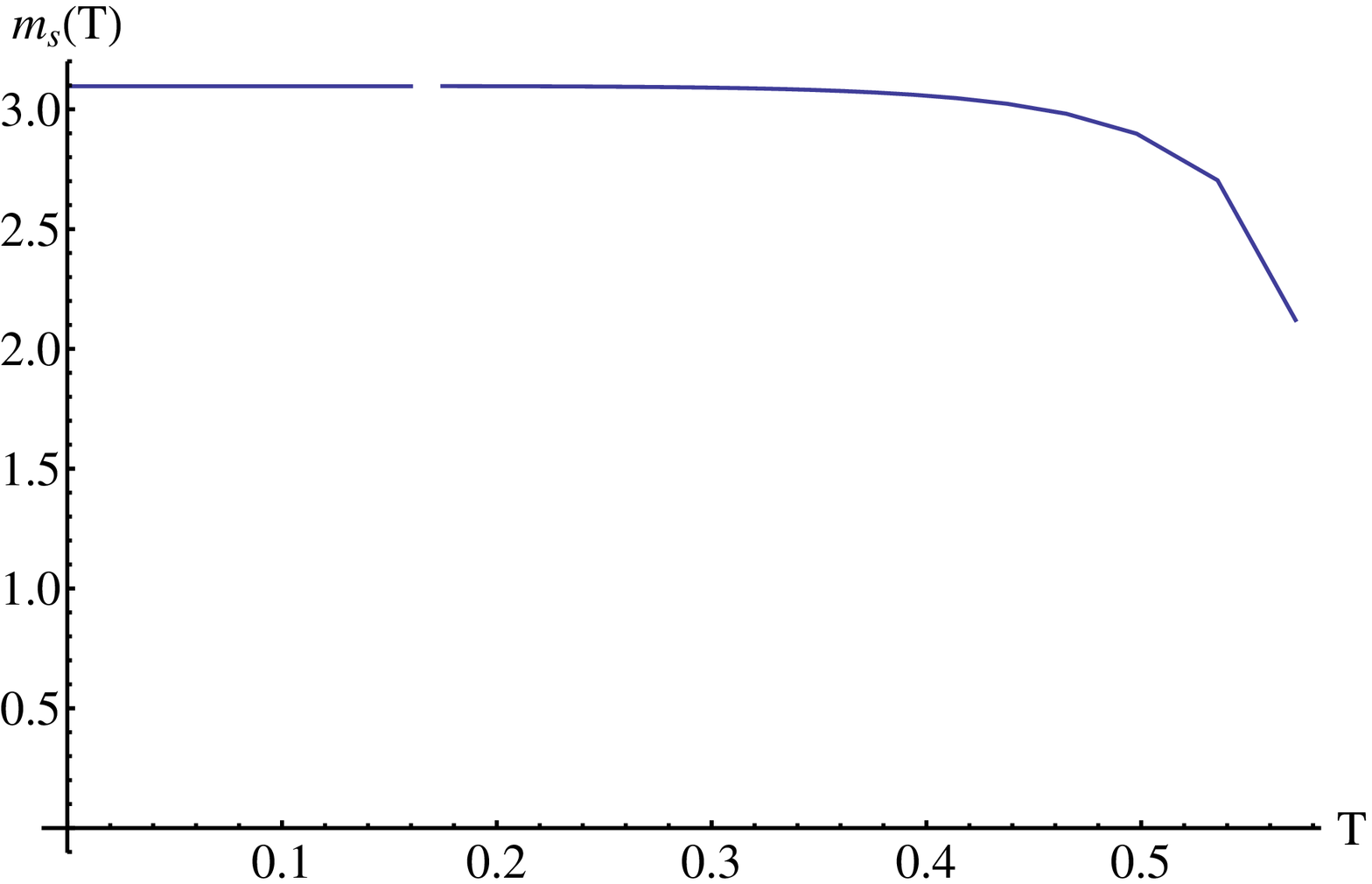}
\includegraphics[width=0.42 \textwidth]{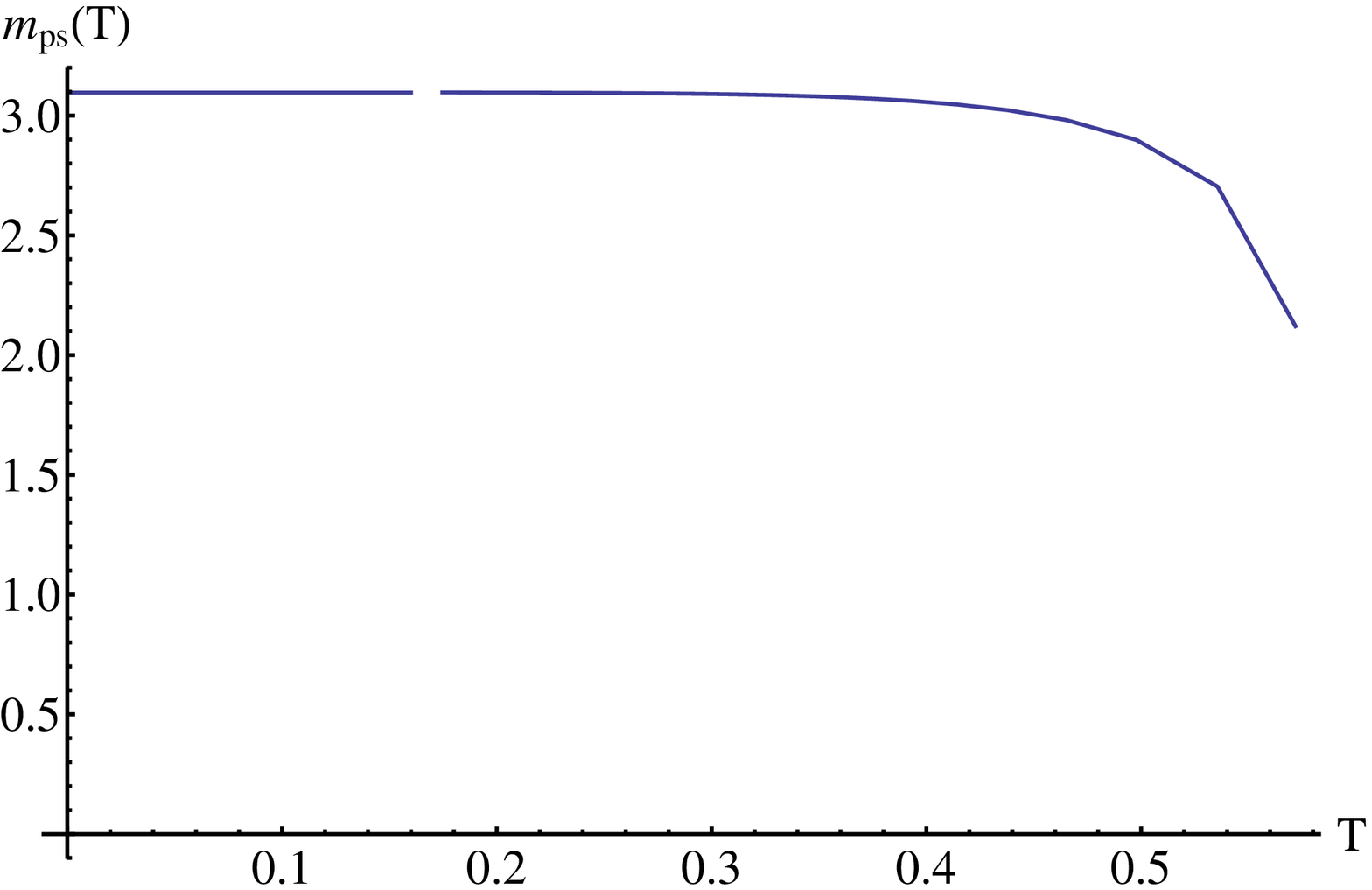}
\end{center}
  \caption{\small The temperature dependent meson masses. Real scalar (left) and pseudo-scalar (right) masses in GeV unit. In confined phase the temperature dependence is trivially constant which is smooth at the deconfinement transition temperature, $T_{dec} \sim$ 160 MeV.
  Note that the disjoint at $T_{dec}$ denotes that we use different backgrounds below and above  $T_{dec}$.
  Above the melting temperature, $T_{fund} \sim$ 571 MeV, there is no discrete spectrum for the meson. }\label{mspsT}
\end{figure}
The fields $\phi, \delta r$ have two linearly independent solutions at $\lambda \rightarrow \infty$ as $\lambda^{0}, \lambda^{-1}$. Here we choose the asymptotic solution as 1/$\lambda$ to have a normalizable mode. Note that
\ba
M_{\phi}^2 &=& \frac{L^3}{U_T} m_{ \phi}^2 = \left(\frac{3}{4\pi}\right)^2 \frac{m_{ \phi}^2}{T^2} = \left(\frac{3}{4\pi}\right)^2 \frac{m_{ \phi}^2}{\bar{M}^2} r_\infty \no
m_{ \phi} &=& \frac{4\pi}{3} \frac{\bar{M}}{\sqrt{r_\infty}} M_{ \phi}.
\ea
The temperature dependent masses are shown in Fig. \ref{mspsT}.
\begin{figure}[h]
\begin{center}
\includegraphics[width=0.31 \textwidth]{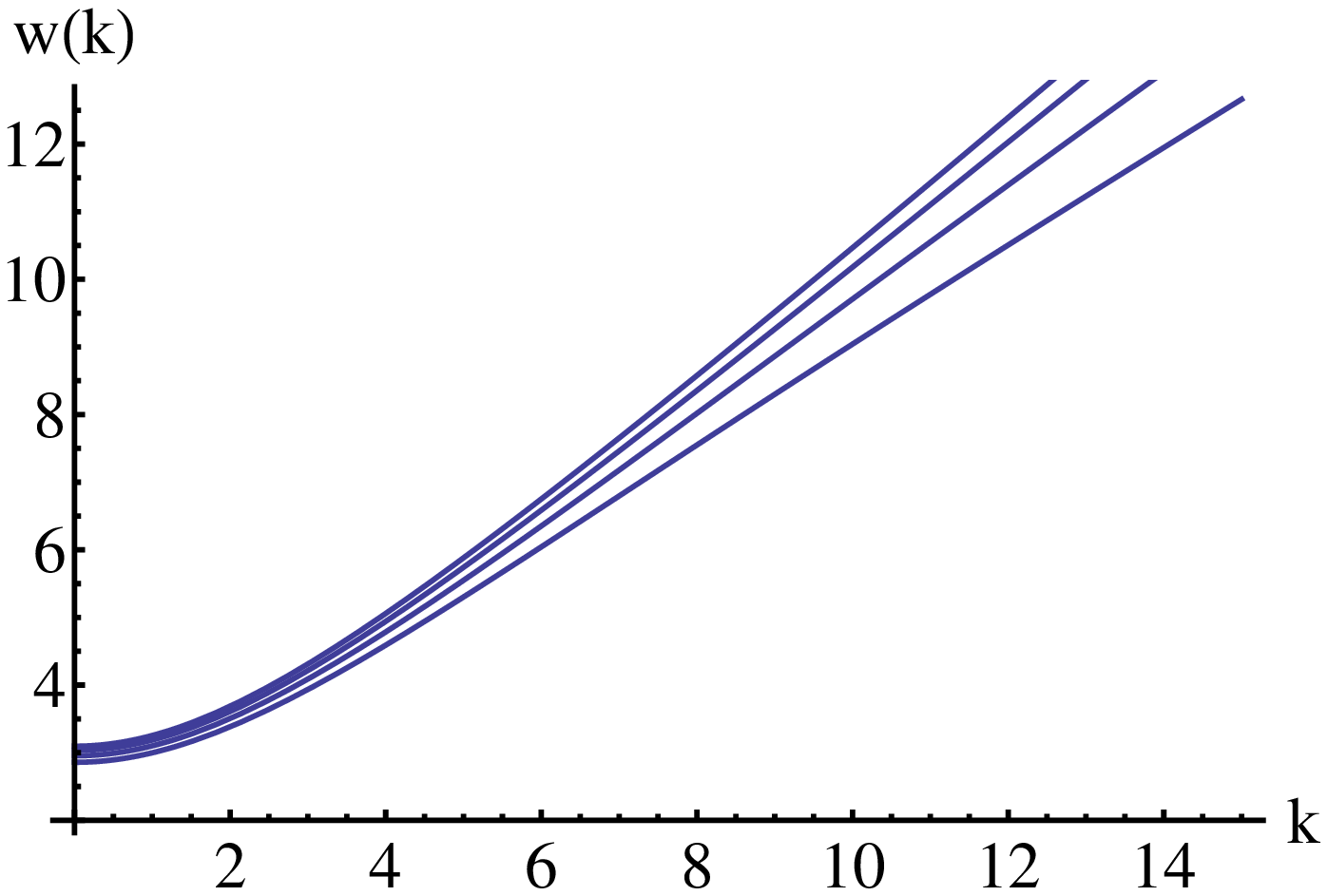}
\includegraphics[width=0.31 \textwidth]{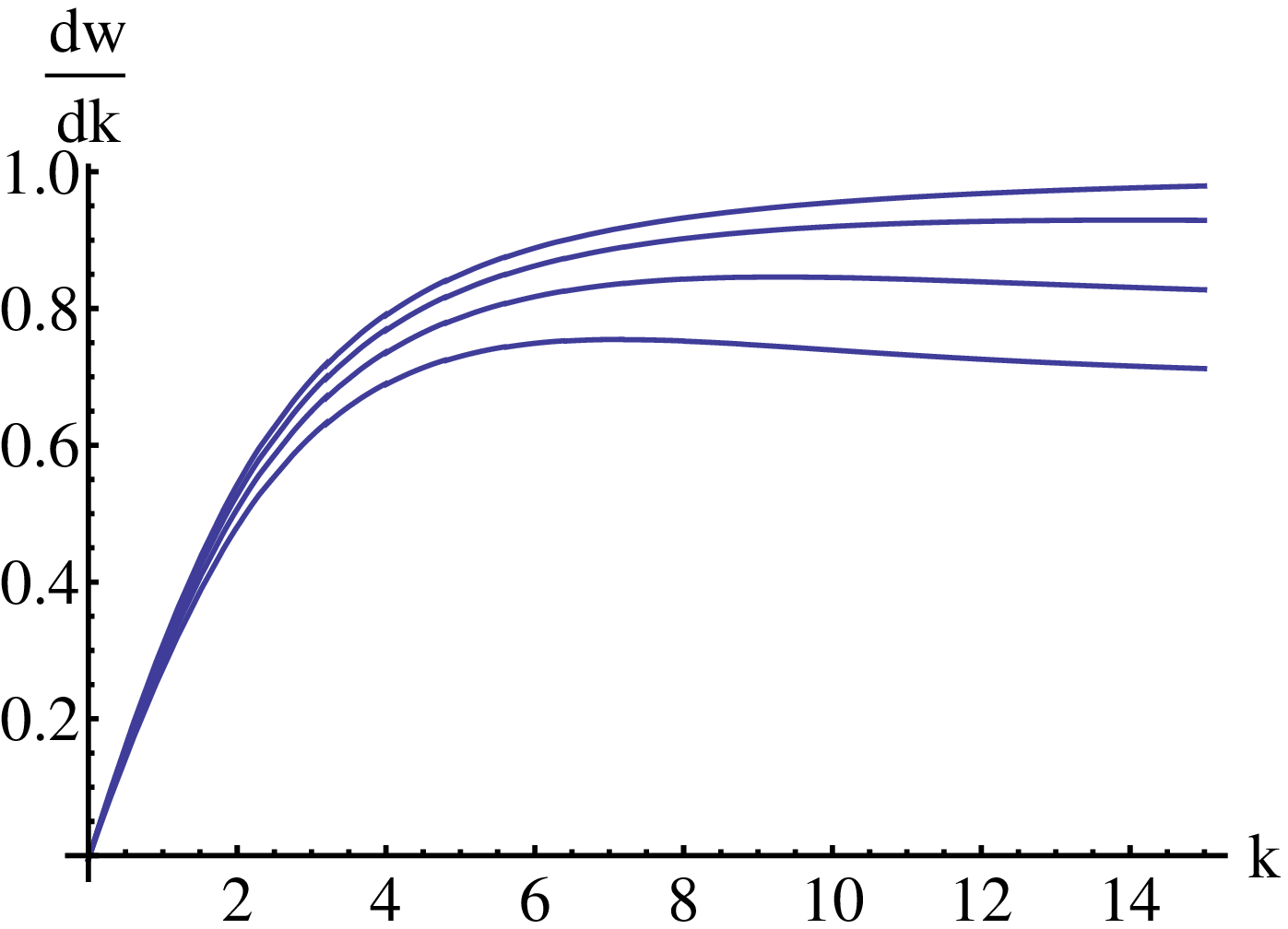}
\includegraphics[width=0.31 \textwidth]{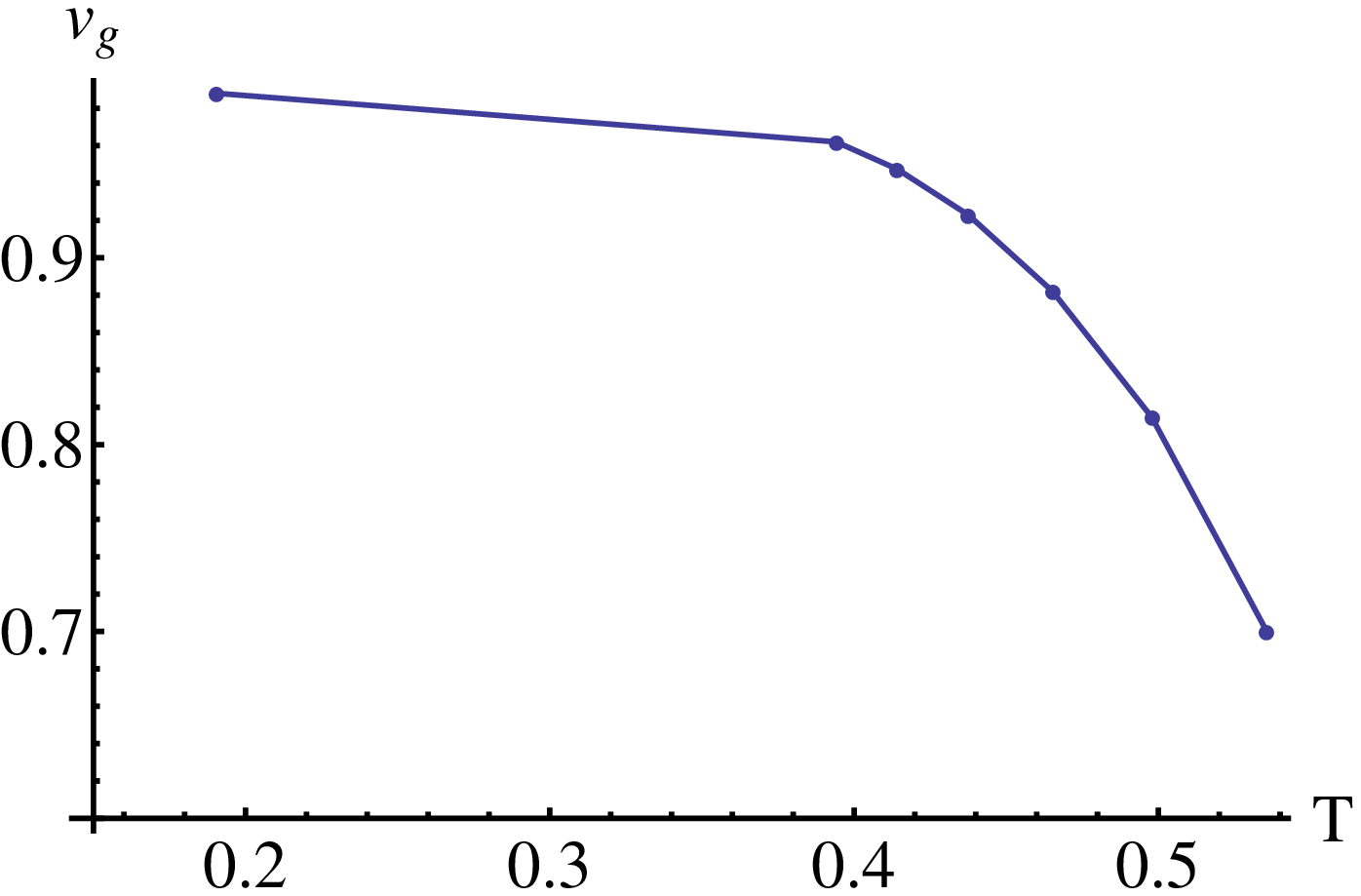}
\end{center}
  \caption{\small Left :
 Dispersion relation for the real scalar with various temperature, T = (1.15, 2.6, 3, 3.24) $T_{dec}$ from top to bottom.
 Middle: $\frac{dw}{dk}$ as a function of k. Right: The group velocity $v_g = \frac{dw}{dk}|_{k \rightarrow \infty}$ as a function of temperature.
 }\label{disS}
\end{figure}
Now we calculate the the dispersion relation of heavy quarkonia, which is important to understand the dissociation and the screening mass
of heavy quarkonium.
The result in Fig. \ref{disS} shows that for sufficiently low temperature but larger than $T_{dec}$,
the group velocity $v_g$ is approaching one, and it is decreasing with increasing temperature.
Note that in D3/D7 model the dispersion relations of mesons was studied~\cite{Ejaz}.

\subsubsection{Gauge field fluctuations}
\begin{figure}
\begin{center}
\includegraphics[width=0.5 \textwidth]{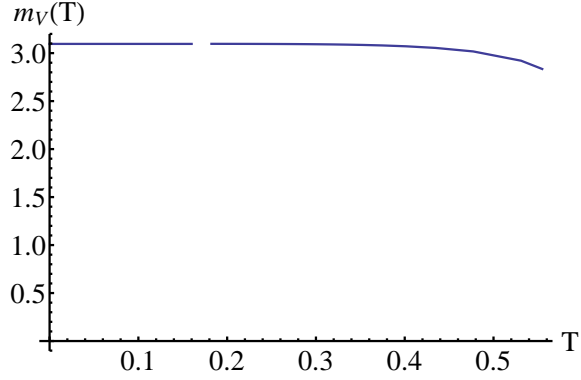}
\end{center}
  \caption{\small The temperature dependent vector meson masses in GeV unit.
  Note that the disjoint at $T_{dec}$ denotes that we use different backgrounds below and above  $T_{dec}$. }
  \label{mVT}
\end{figure}

Now we move on to the gauge field to study vector mass in deconfined phase. The relevant Lagrangian is
\ba \label{FTGFLag1}
\calL_G &=& \frac{\lambda^2}{4} \sqrt{1+\dot{r}^2}\left(1-\left(\frac{U_T^3}{4\rho_v^3}\right)^2 \right)\left(F_{\mu\nu}F^{\mu\nu}
+2F_{\lambda\mu}F^{\lambda\mu}\right) \no
&=&\frac{\lambda^2}{4} \sqrt{1+\dot{r}^2} \left(1-\left(\frac{U_T^3}{4\rho_v^3}\right)^2 \right)\left(g^{\nu\beta}g^{\mu \alpha}F_{\mu\nu}F_{\alpha\beta}
+2g^{\lambda\lambda}g^{\mu \alpha}F_{\lambda\mu}F_{\lambda\alpha}\right)\, .
\ea
{\underline{\textit{Transverse vector}}} \\
The vector fluctuations at finite temperature  are classified into transverse and longitudinal modes due to lack of Lorentz invariance.
 We first consider the transverse part. Assuming that the boundary gauge field propagates only in $x$ direction,
 we obtain the Lagrangian for transverse vector fluctuations as
\be
\calL_G^T = \frac{1}{2} \sqrt{g_0} \bigg[\left(\frac{L}{U}\right)^{3/2} \left(-\frac{1}{f(U)} (\deriv_t A_y)^2 + (\deriv_x A_y)^2\right) + \frac{1}{K (1+\dot{r}_v)} (\deriv_\lambda A_y)^2\bigg].
\ee
The equation of motion for the transverse vector is
\be
\deriv_\lambda\left(\frac{\sqrt{g_0}}{K(1+\dot{r}_v^2)} \deriv_\lambda A_y\right)
+ \sqrt{g_0} \left(\frac{L}{U}\right)^{3/2} \left(-\frac{\deriv_t^2}{f}+\deriv_x^2\right)A_y =0\, ,
\ee
where $\quad \sqrt{g_0}=\lambda^2\sqrt{1+\dot{r}^2} \left[1-\left(\frac{U_T^3}{4\rho_v^3}\right)^2 \right]\left(\frac{L}{U}\right)^{3/2}$.
\begin{figure}[h]
\begin{center}
\includegraphics[width=0.45 \textwidth]{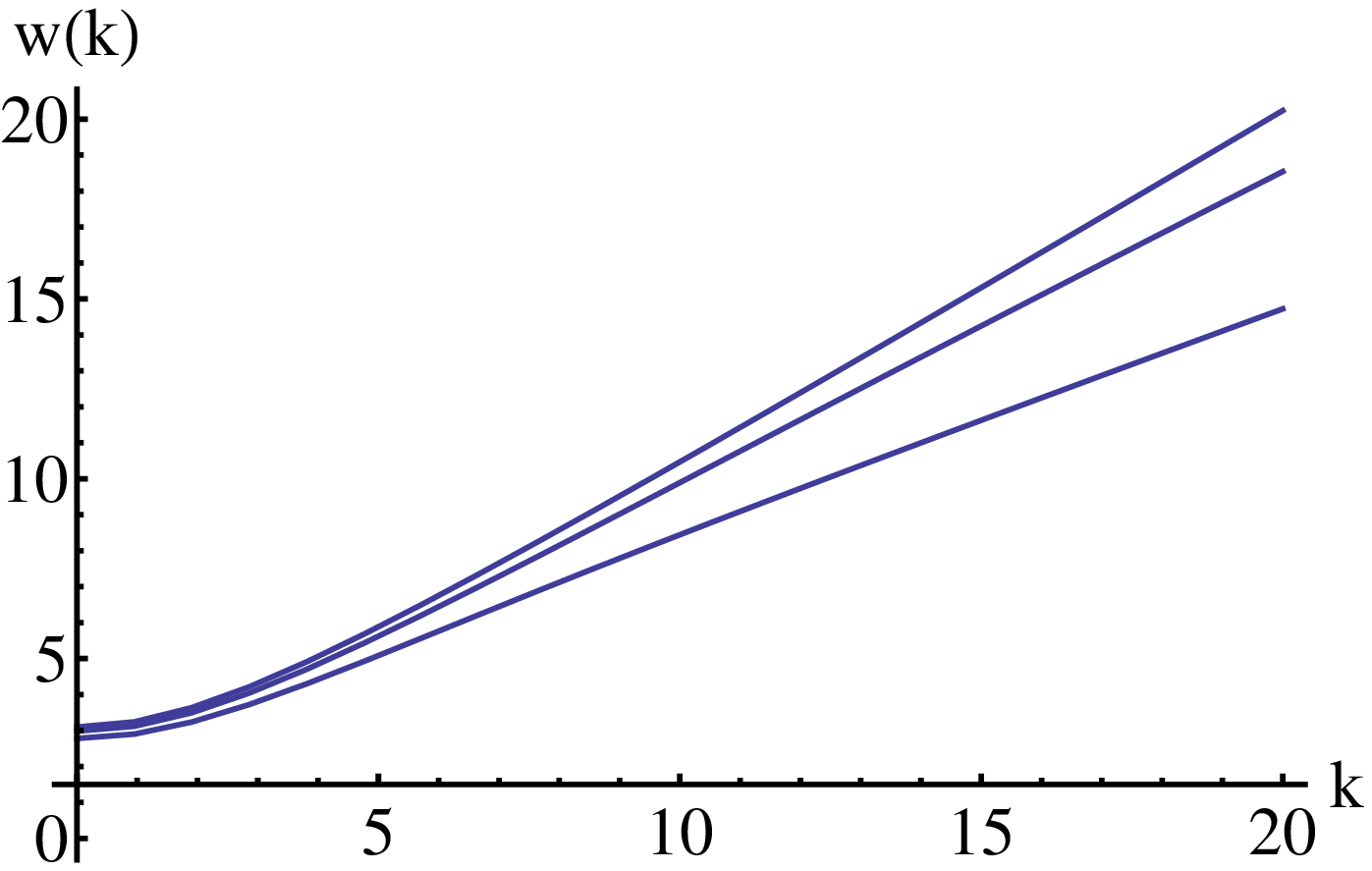}
\includegraphics[width=0.45 \textwidth]{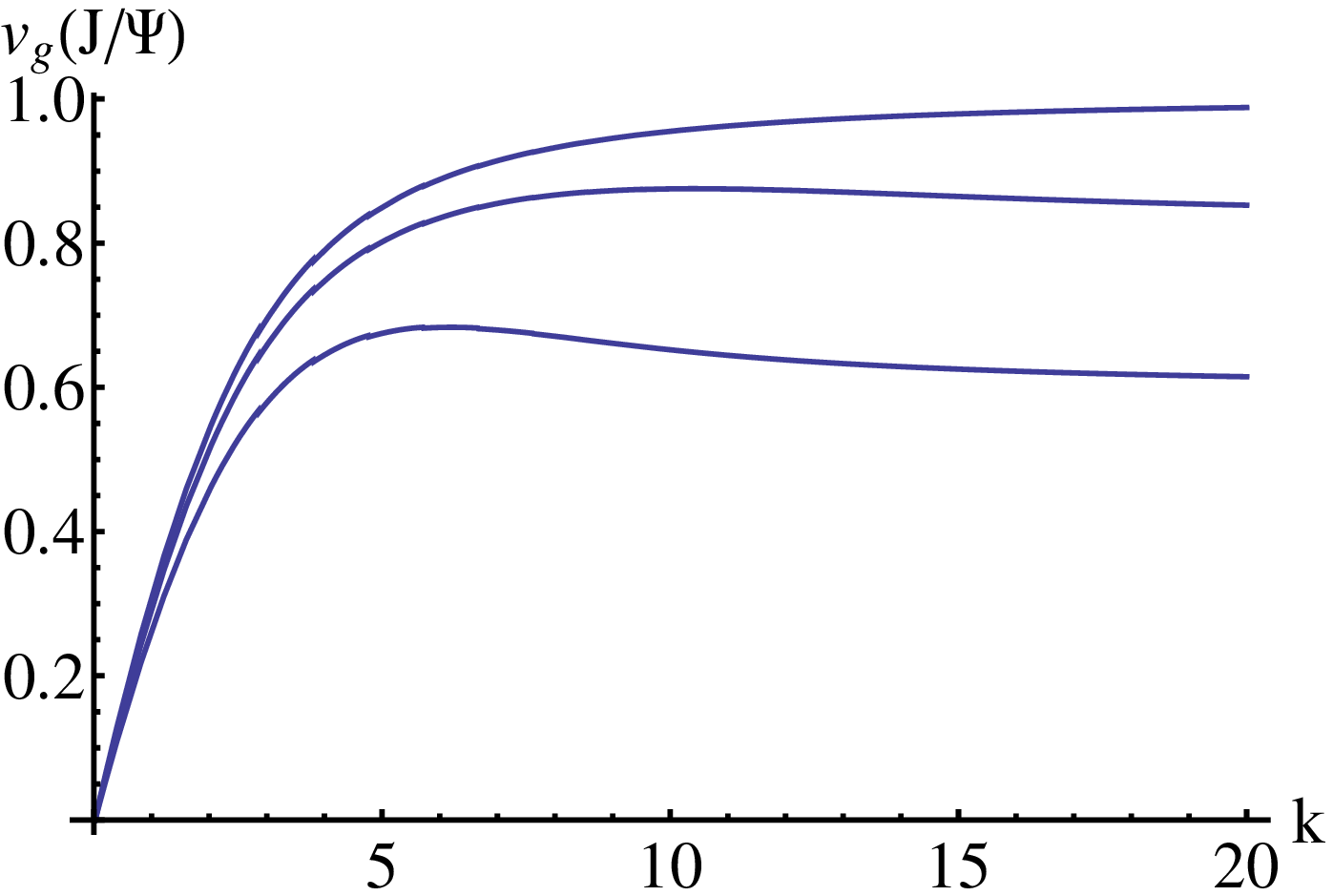}
\end{center}
  \caption{\small
 Dispersion relation for the transverse vector with T = 1.04, 2.91, 3.37 $T_{dec}$ from bottom.}\label{disV}
\end{figure}
We calculate the dispersion relation and group velocity to obtain Fig. \ref{disV}.
Note that $w \sim k^2$ for small k  and $w \sim k$ in large k region.\\
{\underline{\textit{Longitudinal vector}}} \\
We decompose the gauge field into the ortho-normal basis
\be
A_i = \sum_i B_i^{(n)} \psi_n , \quad A_\lambda =\sum_i \varphi^{(n)}\phi_n\,
\ee
to obtain
\ba
\calL_G & = & -\frac{\sqrt{g_0}}{4} \left[\left(\frac{L}{U}\right)^{3/2} \frac{F_{\mu\nu}F_{\alpha\beta}\eta^{\mu\alpha}\eta^{\nu\beta}}{f}+2\frac{(\deriv_\lambda A_i)^2}{K (1+\dot{r}^2)}\right] \no
& = & -\frac{\sqrt{g_0}}{4}\left[\left(\frac{L}{U}\right)^{3/2} \frac{\psi_m \psi_n}{f} F_{\mu\nu}^{(m)} F^{(n)\mu\nu} + 2\frac{\dot{\psi}_m \dot{\psi}_n}{K (1+\dot{r}^2)} B_i^{(m)} B^{(n)i}\right] \, .
\ea
Again, we define the thermal meson mass at zero momentum, $\deriv_i =0$ or $k_i=0$. Then the wave-function of the gauge field satisfies
\be
\int d\lambda  ~ 2 \lambda^2 \left[1-\left(\frac{U_T^3}{4\rho_v^3}\right)^2\right] \left(\frac{L}{U}\right)^{3/2} \frac{\dot{\psi}_m\dot{\psi}_n}{K \sqrt{1+\dot{r}^2}} = m_n^2 \delta_{mn}
\ee
with the normalization condition
\be
\int d\lambda  \frac{\lambda^2}{f} \sqrt{1+\dot{r}^2}  \left[1-\left(\frac{U_T^3}{4\rho_v^3}\right)^2\right] \left(\frac{L}{U}\right)^3\psi_m\psi_n  = \delta_{mn}.
\ee
From these, we obtain the equation of motion for the vector mesons
\be
\deriv_\lambda \left[\left(1-\left(\frac{1}{4\rho_v^3}\right)^2\right) \left(\frac{L}{U}\right)^{3/2}\frac{\lambda^2 }{K \sqrt{1+\dot{r}^2}} \deriv_\lambda \psi_m \right] + \frac{\lambda^2 \sqrt{1+\dot{r}^2}}{U^3} \left(1-\left(\frac{1}{4\rho_v^3}\right)^2\right) \frac{M_V^2}{f}\psi_m =0\, ,
\ee
where $M_{V}^2 = \frac{L^3}{U_T} m_V^2$ is dimensionless.
The meson mass is given by
\be
m_V = \sqrt{\frac{U_T}{L^3}}M_V  .
\ee
We solve the mode equation numerically and show the result in Fig. \ref{mVT}.
We impose $\psi_m\sim 1/\lambda$ at large $\lambda$ due to the normalizability.

As shown in Fig. \ref{mspsT} and Fig. \ref{mVT}, the mass changes smoothly from confining to deconfining phase.
This may be different from  previous studies based
on a bottom-up AdS/QCD model~\cite{KLL07}, QCD sum rule approach~\cite{LeeMorita}, and holographic heavy potential~\cite{KLPS}.
Note, however, that the maximum shift of charmonium mass at $T=1.05 T_c$ is about $200$ MeV~\cite{LeeMorita}, which is roughly $6.7\%$ of the
mass of $J/\psi$ and may be too small to be explained by our approach based on large $N_c$ approximation.

\subsection{Remarks}
Now, we discuss the relevance of the D4/D6 model for light mesons and heavy quarkonia.
One of our basic goals is to study both light mesons and heavy quarkonia on the same footing based on a single D-brane model with a common energy scale $M_{KK}$. Another one is to study heavy quarkonium properties in the high-temperature deconfining phase.
Apart from non-Abelian chiral symmetry, the D4/D6 model might be a good candidate for these purposes since it has both confining and deconfining phases and includes quark mass through embeddings naturally. Especially for a heavy quark system, the D4/D6 model shows some similarity with heavy quarkonium physics.
First, the non-Abelian chiral symmetry is not an issue for heavy quarks.
Second, the D4/D6 model has one more transition at  $T=T_{fund}$ other than the deconfinement transition, which may be associated with the dissociation of heavy quarkonia at high temperature. A positive clue for this is that  $T_{fund}$ is proportional to the square root of a quark, meaning that the charmonium melts relatively at low temperature compared to the bottomonium.
For instance, the dissociation temperature for  a bottomonium $\Upsilon$ is $\sim 2.7 T_c$, while that for a charmonium $J/\psi$
is $\sim 1.3 T_c$~\cite{BSC}.

We finish this section with a summary of the discussion in \cite{Mateos:2007vn} on the usefulness of Dq/Dp systems in studying meson bound states. A Dq/Dp system may be good for $s\bar s$ bound states
 at high temperature since the mesons in the Dq/Dp system are deeply bounded. Even above $T_{fund}$,
  there still exists some broad peak in the spectral function of two-point correlators of mesons.
Therefore $T_{fund}$ is slightly smaller than the dissociation temperature of the heavy quarkonium.
The D4/D6 model may describe some exotic gauge theories. However, there exist certain properties of heavy quarkonia in the quark-gluon plasma that could be understood in the D4/D6 model.

\section{Summary}
We  re-analyzed  the D4/D6 model to study holographic light vector mesons and the properties of heavy quarkonium
in confining and deconfining phases. To treat the light mesons and heavy quarkonium on the same footing, we used the same compactification
scale $M_{KK}$ in both systems.
In confined phase, we observed that the meson spectroscopy of the light meson and heavy quarkonium
 could be described with a single $M_{KK}$.
 We found that like scalar and pseudo-scalar mesons the vector meson mass is
linearly proportional to the square root of the quark mass, when the quark mass is large.
With a $M_{KK}$ fixed by light meson masses, we calculated the mass of heavy quarkonium in confining and deconfining phases.
Unlike previous studies based
on a bottom-up AdS/QCD model~\cite{KLL07}, QCD sum rule approach~\cite{LeeMorita}, and holographic heavy potential~\cite{KLPS},
we found that the mass of heavy quarkonium changes smoothly from confining to deconfining phase.
However, since the maximum shift of charmonium mass at $T=1.05 T_c$ is about $200$ MeV~\cite{LeeMorita}, which is roughly $6.7\%$ of the
mass of $J/\psi$, it is not conclusive if our results are really different from a previous study.
We also obtained the in-medium dispersion relation for heavy quarkonium, which is important to understand the dissociation and the screening mass
of heavy quarkonium.

Certainly there are many things to be improved in our study to be more close to QCD.
We list some of them here.
Surely chiral symmetry should be the first one.
As well known, non-Abelian chiral symmetry is essential to understand light mesons, and it is also
important for heavy-light system due to the light quark.
Second thing is how to include heavy-light meson in this picture with correct chiral symmetry and heavy quark spin symmetry.

\acknowledgments

Y.K. acknowledges the Max Planck Society(MPG), the Korea Ministry of Education, Science and
Technology(MEST), Gyeongsangbuk-Do and Pohang City for the support of the Independent Junior
Research Group at the Asia Pacific Center for Theoretical Physics(APCTP).
The work of SJS  is supported by the National Research Foundation of Korea(NRF)
grant funded by the Korea government(MEST) (R11-2005-021). 
SJS is also supported by the WCU project (R33-2008-000-10087-0) and   by
NRF  Grant R01-2007-000-10214-0.


\end{document}